\documentstyle[pra,aps,twocolumn]{revtex}

\begin{document}

\title{The temperature of the quantum vacuum}

\author{Adonai S. Sant'Anna \and Daniel C. Freitas}

\address{Department of Mathematics, Federal University at Paran\'a\\P.O. Box 19081, Curitiba, PR, 81531-990, Brazil.}


\maketitle

\begin{abstract}
We propose that the quantum vacuum may be considered as a gas of virtual photons which carry a non-vanishing linear momentum as well as a non-vanishing energy. We study, in particular, the Casimir effect in order to show that these virtual photons should satisfy a Fermi-Dirac statistics, which implies a non-zero temperature of the vacuum state.\\
\end{abstract}

\draft{PACS number(s): 05.30.-d, 12.20.-m}

\section{Introduction}

	We need to recall some important aspects of quantum electrodynamics (QED) in order to present a theory of a quantum gas associated to the vacuum state. Here follows some very brief remarks on the quantum vacuum and the Casimir effect:

\begin{enumerate}

\item According to QED, the vacuum state carries an energy $\frac{1}{2}\hbar\omega$ and a linear momentum $\frac{1}{2}\hbar {\bf k}$, for each field mode (e.g., a material oscillator), where $\hbar$ is the reduced Planck's constant, $\omega$ is the frequency of the field mode and ${\bf k}$ is the wave vector. This vacuum energy is usually refered to as the zero-point energy.

\item In 1948 H. Casimir \cite{Casimir-48} showed one intriguing consequence of the zero-point energy: there is an attractive force between two perfectly conducting parallel plates, standing face to face in vacuum at a distance $d$ much smaller than their lateral extensions. This effect is now known as the Casimir effect.

\item In 1988, P. W. Milonni, R. J. Cook and M. E. Goggin \cite{Milonni-88} proposed a physical interpretation of the Casimir effect in terms of radiation pressure from the vacuum. According to these authors, ``the virtual photons of the vacuum carry linear momentum $\frac{1}{2}\hbar{\bf k}$''. Notwithstanding, Milonni {\em et al\/} did not discuss any statistical aspect concerning these virtual photons.

\item In 1991 G. Barton \cite{Barton-91a,Barton-91b} used standard statistical and quantum physics to analyze the fluctuations of the Casimir stress exerted on a flat perfect conductor
by the vacuum electromagnetic fields in adjacent space. According to Barton \cite{Barton-91a} ``though a simple photon-gas model of the Casimir effect, like that suggested by Milonni {\em et al\/} \cite{Milonni-88}, is suited to interpreting the mean Casimir force, it affords no immediate insight into its fluctuations.'' We agree with G. Barton. Nevertheless, we consider that a statistical picture of the quantum gas of virtual photons of the vacuum may lead to the calculation of its temperature.

\item In 1997 S. K. Lamoreaux \cite{Lamoreaux-97} performed an experiment which demonstrates the Casimir force in the 0.6 to 6 $\mu$m range, within a 5\% error.

\end{enumerate}

	In this paper we propose a statistical theory of the virtual photons of the vacuum that is in agreement with experimental data. As a consequence of our theory, we can infer a non-zero temperature of the quantum vacuum. By quantum vacuum we mean the vacuum state in the context of QED.

\section{The Vacuum as a Gas of Virtual Photons}

	We focus on the simplest case of the Casimir force: the two plates system, as described above. We propose that the zero-point energy of the quantum vacuum may be associated to virtual photons of a quantum mechanical ideal gas.

	It is well known that the boundary condition for the field modes appropriate to the interior of a rectangular parallelepiped of sides $L_x = L_y = L$ and $L_z = d$ are given by

\begin{equation}
\frac{k_x}{\pi} = \frac{l}{L},\;\;\frac{k_y}{\pi} = \frac{m}{L},\;\;\frac{k_z}{\pi} = \frac{n}{d}\label{boundary}
\end{equation}

\noindent
where $d$ is the distance between the two perfectly conducting plates standing face to face; $l$, $m$, and $n$ are positive integer values and zero; $\pi = 3.14159$; and $k_{xyz}$ are components of the vector ${\bf k}$.

	The expected number of photons that strike the area $dS$ of one of the plates within the time interval $dt$ is

\begin{equation}
f(k)\frac{1}{\pi^{3}}dk_{x}dk_{y}dk_{z}\cos \gamma\, c\,dt\,dS,
\end{equation}

\noindent
where $\cos\gamma = k_z/k$, $c$ is the speed of light in the vacuum, and $f(k)$ is the distribution function of the virtual photons in the vacuum. The factor $\frac{1}{\pi^3}$ is justified by the boundary condition given in equation (\ref{boundary}). The factor $\cos \gamma\, c\,dt\,dS$ is an element of volume $dV$.

	The momentum delivered to the plate by a single reflected photon is equal to the negative of the change in the momentum of the photon. In other words, the momentum is equal to $2\frac{1}{2}\hbar k_{z}$, if we consider the plate perpendicular to the $z$ component of the $xyz$ system of coordinates. Therefore, the expected linear momentum transfered to an area $dS$ on the plate during the time interval $dt$ is

\begin{equation}
\frac{\hbar}{\pi^{3}}\frac{k_{z}^{2}}{k}f(k)dk_{x}dk_{y}dk_{z}c\,dt\,dS.\label{momentum}
\end{equation}

	The force on the plate is obtained by dividing (\ref{momentum}) by $dt$. The pressure is obtained by dividing the force by $dS$. We denote the inward pressure as $P_{in}$ and the
outward pressure as $P_{out}$. Hence:

\begin{equation}
P_{in} = \frac{\hbar
c}{\pi^{3}}\int_{0}^{\infty}dk_{x}\int_{0}^{\infty}dk_{y}\int_{0}^{\infty}dk_{z}\frac{f(k)k_{z}^{2}}{\sqrt{k_{x}^{2}+k_{y}^{2}+k_{z}^{2}}}.
\end{equation}

\noindent
and

\begin{equation}
P_{out} = \frac{\hbar
c}{\pi^{2}d}\sum_{n=1}^{\infty}\int_{0}^{\infty}dk_{x}\int_{0}^{\infty}dk_{y}\frac{f(k)(\frac{n\pi}{d})^{2}}{\sqrt{k_{x}^{2}+k_{y}^{2}+(\frac{n\pi}{d})^{2}}},\label{Pout}
\end{equation}

\noindent
where $k = \sqrt{k_x^2+k_y^2+k_z^2}$. The summation in equation (\ref{Pout}) is due to the fact that the continuum approximation is not valid between the plates, since we are assuming $d\ll L$.

	After some simple algebra, we can write the difference $P_{out} - P_{in}$ as:

\begin{eqnarray}
P_{out} - P_{in} = 
\frac{\pi^{2}\hbar c}{4d^{4}}\left[\sum_{n=1}^{\infty}n^{2} \int_{0}^{\infty}dx\frac{f(\sqrt{x+n^{2}})}{\sqrt{x+n^{2}}} -\right.\nonumber\\
\left.-\int_{0}^{\infty}du\,u^{2}\int_{0}^{\infty}dx\frac{f(\sqrt{x+u^{2}})}{\sqrt{x+u^{2}}}\right],\label{diff}
\end{eqnarray}
where, by change of variables, $f(k) = f(\sqrt{x+u^{2}})$, $x = x'^{2}
= \frac{k_{x}^{2}d^{2}}{\pi^{2}}+\frac{k_{y}^{2}d^{2}}{\pi^{2}}$, $u =
k_{z}\frac{d}{\pi}$, $\theta = \tan\left(\frac{k_{y}}{k_{x}}\right)$,
and $dk_{x}dk_{y} = x'dx'd\theta\frac{\pi^{2}}{d^{2}}$. Equation (\ref{diff}) is well known in the literature \cite{Milonni-94}, although it is usually derived from other assumptions. The expression
$f(\sqrt{x+u^{2}})$ in equation (\ref{diff}), for example, is usually interpreted as a cutoff function such that $f(k) = 1$ if $k\ll k_c$ and $f(k) = 0$ if $k\gg k_c$, where $k_c$ is a constant. It is usually argued that the assumption of perfectly conducting walls breaks down at small wavelengths like $k_c = \frac{1}{a_0}$, where $a_0$ is the Bohr radius. When these properties of a cutoff function are assumed, it can be proved, by using the Euler-MacLaurin summation formula, that the series given by the equation above converges to:

\begin{equation}
P_{out} - P_{in} = -\frac{\pi^{2}\hbar c}{240d^{4}}.\label{Casimir}
\end{equation}

	The Euler-MacLaurin summation formula is given by:

\begin{eqnarray}
\sum_{n=1}^\infty F(n) - \int_0^\infty dk F(k) =\nonumber\\
 -\frac{1}{2}F(0)+\sum _{r=1}^{\infty }(-1)^{r-1}\frac{B_{r}}{(2r)!}F^{(2r-1)}(0)
\end{eqnarray}

\noindent
for $F(\infty)\to 0$, where $B_{r}$ is the $r$-th Bernoulli's coefficient.

	It is important to recall that equation (\ref{Casimir}) is in agreement with the usual theory of the Casimir effect \cite{Milonni-94}, as well as with experimental data \cite{Lamoreaux-97}.

\section{The Distribution Function}

	The natural question is: what about $f(k)$? Is $f(k)$ a Maxwell-Boltzmann, Bose-Einstein,  Fermi-Dirac or any other sort of parastatistic distribution function? Let us see the possibilities:

\begin{description}

\item[Maxwell-Boltzmann] From the physical point of view, Maxwell-Boltzmann is nonsense, since we are talking about QED and {\em indistinguishable\/} particles. From the mathematical point of view, there is no choice of parameters in Maxwell-Boltzmann statistics which corresponds to the mathematical properties of a cutoff function, as described in the Section above. In other words, if we assume $f(k)$ as a Maxwell-Boltzmann distribution function, there is no way to prove that the series given in equation (\ref{diff}) converges to the expression given by equation (\ref{Casimir}).

\item[Bose-Einstein] From the physical point of view, Bose-Einstein is nonsense, although  we are talking about virtual {\em photons\/}. The reason is quite simple: we assume just {\em one\/} virtual photon for each field mode. Such a restriction sounds like Pauli's Exclusion Principle. From the mathematical point of view, there is no choice of parameters in Bose-Einstein statistics which corresponds to the mathematical properties of a cutoff function. This mathematical argument is similar to the case of Maxwell-Boltzmann distribution function.

\item[Fermi-Dirac] Fermi-Dirac statistics is sound in the context of our problem. From the physical point of view, we consider just {\em one\/} virtual photon for each field mode, if  $k\ll k_c$, and {\em zero\/} photons if $k\gg k_c$. From the mathematical point of view, we can easily define parameters in Fermi-Dirac statistics in order to obtain the same properties of a cutoff function.

\item[Parastatistics] We do not consider that it is necessary to study the possibility of a parastatistic distribution \cite{Haag-92} of virtual photons, since Fermi-Dirac works well.

\end{description}

\section{The Temperature of the Vacuum}

	If we assume that $f(k)$ is a Fermi-Dirac distribution function, then:

\begin{equation}
f(k)=\frac{1}{e^{\alpha +\beta k}+1}\label{FD}
\end{equation}

\noindent
where $\alpha$ is a normalization constant usually refered to as the {\em affinity\/}, and 

\begin{equation}
\beta = \frac{1}{KT},\label{beta}
\end{equation}

\noindent
where $K$ is the Boltzmann constant and $T$ is the absolute temperature.

	We can rewrite equation (\ref{FD}) as:

\begin{equation}
f(k)=\frac{1}{e^{\beta(k-k_{c})}+1}
\end{equation}

\noindent
where  $k_{c}=-\frac{\alpha}{\beta}$ is the only value such that the second derivative of $f(k)$ vanishes, i.e., $f''(k) = 0$. This last condition allows us to determine the value of $k_c$, as a function of the absolute temperature.

	It is obvious, from equations (\ref{FD}) and (\ref{beta}), that $T\neq 0$. The question, now, is: how to calculate $T$? If we assume that $k_c = 1/a_0$, where $a_0$ is the Bohr radius, we have $\omega_c = k_c c = 5.666\times 10^{18} s^{-1}$. Therefore, $T = -1.369\times 10^{33}\times\alpha^{-1}K$, since $\beta = -\alpha/k_c = 1/(KT)$. The constant $\alpha$ is a negative real number. So, the temperature $T$ depends on the total number of virtual photons which are interacting with the plates.

	It is well known that the quantum vacuum does have linear momentum, energy, fluctuating electromagnetic fields, and radiation pressure. Now, we are proving that it does have temperature. This temperature depends on the existence of boundary conditions, that is, it depends on the presence of matter.

\section{Remarks}

\begin{itemize}

\item The non-zero temperature of the quantum vacuum state is not surprising, since we are talking about a {\em gas\/} of virtual photons. Nevertheless, it is well known that a dipole oscillating with a given frequency, moving with velocity $v$ through a thermal field, experiences a frictional force $F$ which is opposite to $v$ \cite{Milonni-94}. But in our case, the force $F$ is zero, since it depends on the spectral energy  density and not on the temperature. Our work is in agreement with the fact that the spectral energy density of the vacuum is the same in all inertial frames.

\item The Fermi-Dirac distribution function for virtual photons does not violate the spin-statistcs theorem \cite{Ryder-94}, since such a theorem is valid just for quantum fields defined by the creation operator on the vacuum state. In this paper we are dealing directly with the zero-point energy state. So, our work is out of the scope of the spin-statistics theorem.

\item As a final remark, we say that this work is inspired in \cite{Suppes-96}, where P. Suppes, A. S. Sant'Anna and J. A. de Barros presented a particular and classical theory of the Casimir effect. In this paper we develop similar ideas in the context of QED and we explicitely discuss the statistical distribution of the virtual photons.

\end{itemize}

\end{document}